\documentclass[sigconf,authorversion]{acmart}

\AtBeginDocument{%
  \providecommand\BibTeX{{%
    \normalfont B\kern-0.5em{\scshape i\kern-0.25em b}\kern-0.8em\TeX}}}

\copyrightyear{2021} 
\acmYear{2021} 
\setcopyright{acmlicensed}\acmConference[CHI '21]{CHI Conference on Human Factors in Computing Systems}{May 8--13, 2021}{Yokohama, Japan}
\acmBooktitle{CHI Conference on Human Factors in Computing Systems (CHI '21), May 8--13, 2021, Yokohama, Japan}
\acmPrice{15.00}
\acmDOI{10.1145/3411764.3445299}
\acmISBN{978-1-4503-8096-6/21/05}

\begin{document}

\title[Creepy Technology]{Creepy Technology:\\What Is It and How Do You Measure It?}


\author{Pawe\l{} W. Wo\'{z}niak}
\orcid{0000-0003-3670-1813}
\affiliation{
\institution{Utrecht University}
\city{Utrecht}
\country{the Netherlands}}
\email{p.w.wozniak@uu.nl}

\author{Jakob Karolus}
\affiliation{
  \institution{LMU Munich}
  \city{Munich}
  \country{Germany}
}
\email{jakob.karolus@ifi.lmu.de}

\author{Florian Lang}
\affiliation{%
  \institution{LMU Munich}
  \city{Munich}
  \country{Germany}
}
\email{florian.lang@ifi.lmu.de}

\author{Caroline Eckherth}
\affiliation{%
  \institution{LMU Munich}
  \city{Munich}
  \country{Germany}
}
\email{caroline.eckerth@gmx.de}

\author{Johannes Sch\"{o}ning}
\affiliation{%
  \institution{University of Bremen}
  \city{Bremen}
  \country{Germany}
}
\email{schoening@uni-bremen.de}
\author{Yvonne Rogers}

\affiliation{%
  \institution{University College London}
  \city{London}
  \country{United Kingdom}
}
\affiliation{
    \institution{University of Bremen}
  \city{Bremen}
  \country{Germany}
}
\email{y.rogers@ucl.ac.uk}

\author{Jasmin Niess}
\affiliation{%
  \institution{University of Bremen}
  \city{Bremen}
  \country{Germany}
}
\email{niessj@uni-bremen.de}


\renewcommand{\shortauthors}{Pawe\l~W. Wo\'{z}niak et al.}

\newcommand{\todop}[1]{\textsf{\textbf{\textcolor{green!55!blue}{[Pawel: #1]}}}}
\newcommand{\todojas}[1]{\textsf{\textbf{\textcolor{yellow!55!red}{[Jasmin: #1]}}}}
\newcommand{\todojak}[1]{\textsf{\textbf{\textcolor{green!55!red}{[Jakob: #1]}}}}
\newcommand{\todof}[1]{\textsf{\textbf{\textcolor{red!55!blue}{[Florian: #1]}}}}
\newcommand{\scale}{Perceived Creepiness of Technology Scale}
\newcommand{\scaleshort}{PCTS}

\makeatletter
\newcommand\footnoteref[1]{\protected@xdef\@thefnmark{\ref{#1}}\@footnotemark}
\makeatother

\begin{abstract}
Interactive technologies are getting closer to our bodies and permeate the infrastructure of our homes. While such technologies offer many benefits, they can also cause an initial feeling of unease in users. It is important for Human-Computer Interaction to manage first impressions and avoid designing technologies that appear creepy. To that end, we developed the Perceived Creepiness of Technology Scale (PCTS), which measures how creepy a technology appears to a user in an initial encounter with a new artefact. The scale was developed based on past work on creepiness and a set of ten focus groups conducted with users from diverse backgrounds. We followed a structured process of analytically developing and validating the scale. The PCTS is designed to enable designers and researchers to quickly compare interactive technologies and ensure that they do not design technologies that produce initial feelings of creepiness in users.
\end{abstract}




\maketitle

\section{Introduction}
When creating novel interactive technologies, be it for research or practical purposes, managing first impressions is key~\cite{harrison_infographic_2015,nourani_investigating_2020,reinecke_predicting_2013}. A technology that looks intimidating, scary or unpleasant is unlikely to engage the user's willingness to interact with it. This challenge becomes even more salient when dealing with technologies that reflect recent trends in Human-Computer Interaction (HCI) such as wearable computing~\cite{kelly_wear_2016} or sensory amplification~\cite{schmidt_augmenting_2017}. While such trends promise attractive technological futures, they also envision many technologies that could initially be perceived negatively. As a consequence, designers of future technologies need to understand how to build technologies that offer positive first impressions. This gives rise to the need for methods that would enable designers to compare alternative prototypes in terms of first impressions.

The HCI field has a history of studying technologies that users perceive as potentially creepy. Privacy research used the term \emph{creepy} to describe technologies that were perceived as potentially encroaching on the users' privacy, e.g.~\cite{pierce_smart_2019}. Research in avatars and robots equated \emph{creepy} with \emph{uncanny}~\cite{ho_human_2008} to denote humanoid representations that produce a certain unease in users. However, the use of the term was not exclusive to these two domains. Past research has reported that self-driving cars~\cite{moore_case_2019}, traffic lights~\cite{paredes_fiat-lux_2016}, headphones~\cite{grinter_ears_2002}, voice assistants~\cite{seymour_does_2020} or toilets~\cite{QuantifiedToilet} could also be creepy. 

More and more people are beginning to encounter creepy technologies in their everyday lives. Newspapers report about technologies from dating apps~\cite{burgess_spotify_2018} to speakers~\cite{dugan_facebooks_2018} being perceived as creepy. A recent concern is facial recognition technology, which is beginning to be deployed in physical and online stores to help retailers improve how they customise the shopping experience. Smart AI platforms are emerging that can detect at a glance our gender, race, approximate age, where and how long we have been looking at something, and in what emotional state we are~\cite{hill_secretive_2020}. Further, creepy technologies are not only developed by corporations but also by users themselves. Modern development tools enable proficient users to develop personal apps such as a chatbot for talking to a late friend based on past conversations~\cite{matei_new_2017}. In this case, even the creator of the chatbot was concerned that the application could be creepy. As users are increasingly likely to experience creepiness in everyday interactions, HCI needs to understand more about the phenomenon to minimise the negative impact of future interactive technologies.

\begin{figure*}
    \centering
    \includegraphics[width=\textwidth]{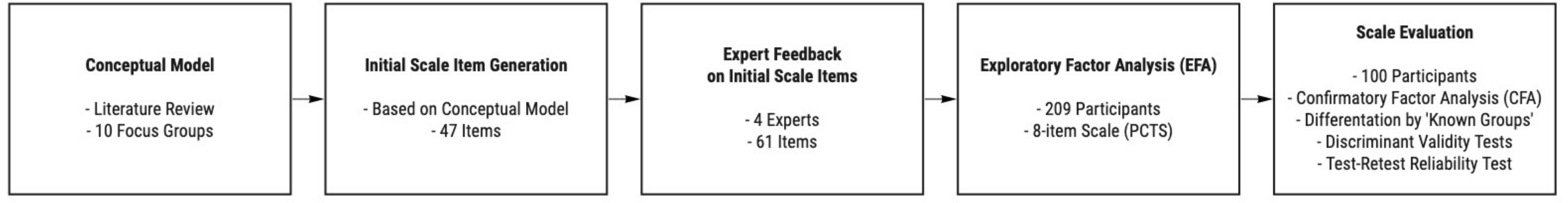}
    \caption{The scale development process which we followed in this paper. The workflow is a selection of development and evaluation methods suggested by Boateng et al.~\protect\cite{boateng_best_2018}.}
    \label{fig:flow}
\end{figure*}

To this end, this paper explores the qualities in technology that give users the heebie-jeebies. We propose a structured framing of the \emph{creepiness} of interactive technologies and proposes an accompanying measurement instrument---the \scale{} (\scaleshort{}). We followed a structured scale development process where we first formed a conceptual understanding of creepiness, followed by empirically building the scale using the guidelines collected by Boateng et al.~\cite{boateng_best_2018}. We first investigated past work in HCI and identified papers which reported technologies being creepy outside of the humanoid or robotics fields. To empirically explore how users think about creepiness, we then conducted a set of ten focus groups where users expressed their opinions about potentially creepy technologies. Based on the literature and our analysis of the focus group content, we then proposed a general framing of the concept for HCI. The elements of the model served as an inspiration to generate initial items for the scale, which were then subjected to an expert review. We used exploratory factor analysis to reduce the number of items and obtain the final scale, which was then validated in a number of evaluation assessments. For an overview of our scale development process, see Figure~\ref{fig:flow}. Our work offers the first, to our knowledge, conceptualisation of creepiness in HCI and a validated scale for assessing the creepiness of interactive artefacts.

\section{Related Work}
To frame our inquiry, we first chart the use of the concept of creepiness and the adjective \emph{creepy} in past research. We report on how the terms were used in privacy research and Human-Robot Interaction. We then investigate the relationship between creepiness and acceptability. Finally, we report on how creepiness was ascribed to non-humanoid digital artefacts in past HCI research.

\subsection{Creepiness in privacy research}
Studies in privacy of personal technologies have extensively used the term creepy to refer to technologies that are perceived as potentially threatening privacy. Creepiness was particularly ascribed to technologies that collect personal information about their users. Pierce~\cite{pierce_smart_2019} researched speculative scenarios for home cameras and concluded that minute details in the design of home devices led to different levels of creepiness. This work calls for unpacking the reasons behind creepiness and shows how creepy a device is perceived is influenced by diverse factors. Zhang et al.~\cite{zhang_creepy_2014} studied how targeted advertisements evoked feelings of creepiness in users. The study focused on the consequences of this on the use of social media and not the sources of creepiness \textit{per se}. Also, in the context of advertising, Ur et al.~\cite{ur_smart_2012} reported that \emph{creepiness} was associated with the feeling of being followed. Importantly for our understanding of creepiness, Shklovski et al.~\cite{shklovski_leakiness_2014} found that creepiness was not connected to an anticipated negative end result of using a technology while studying mobile app use. In this context, both Sklovski et al. and Phelan et al.~\cite{phelan_its_2016} equated \emph{creepy} with \emph{disturbing}. The later paper underlined the intuitiveness of the concept with creepiness (\emph{intuitive concern}) being less rational than \emph{considered concern}. Other research also reported creepiness in the context of privacy violation when users were involved in unsolicited meetings on social media~\cite{barkhuus_student_2010} and crowdworking~\cite{hanson_taking_2020}. The examples listed here are just a few, illustrating the breadth of the use of the term \emph{creepy} in privacy research. While this body of research addresses a broad scope of applications, it shares a common understanding of creepiness as an, often unspoken and innate, anticipation of the technology violating ethical principles held by the user. Our work is inspired by accounts of creepiness in privacy research. We aim, however, to broaden the scope of understanding creepiness beyond privacy concerns.

\subsection{Creepiness of humanoid avatars and robots}
In research on virtual avatars and Human-Robot Interaction (HRI) the notion of creepiness is primarily associated with `uncanniness' and the uncanny valley phenomenon, i.e. unsettling feelings experienced by someone elicited by an artefact's spooky resemblance to a human being or other animate beings. Schwind et al.~\cite{schwind_wheres_2017} reported how some representations of the users' hands evoked feelings of creepiness. Early HRI research showed that creepiness was prevalent when robots offered emergency help~\cite{murphy_robot-assisted_2004}. Lin et al.~\cite{lin_parental_2020} reported that parents wanted to explicitly limit the creepiness of robots when they allowed them to interact with their children. The representation of faces for both robots~\cite{kalegina_characterizing_2018} and virtual avatars~\cite{mcdonnell_face_2010} elicited feelings of creepiness related to the mismatch between their appearance and the user's expectations. L\"{o}ffler et al.~\cite{loffler_uncanny_2020} proposed a different interpretation of creepiness. They used a scale where \emph{creepy} was the opposite of \emph{friendly} to assess the perception of animal-like robots. Creepiness has also been viewed as a pragmatic concern, lowering the effectiveness of interaction when robotic assistants helped with analytical tasks~\cite{somanath_integrating_2013}.
HRI work has contributed a number of understandings of creepiness and identified technologies being potentially creepy as a key concern when building new interactive artefacts. In this paper, we extend the notion of creepiness based on HRI experiences and broaden the scope of potentially creepy technologies beyond robots and human-like artefacts.

\subsection{Acceptability and creepiness in HCI}
Previous research in HCI has often associated the term creepy with social unacceptability. For example, the WEAR scale~\cite{kelly_wear_2016} explicitly used the adjective \emph{creepy} as a contribution to the acceptability scale. Consequently, it might appear that creepiness is a subordinate concept to acceptability. This, however, is in conflict with past work discussed above, which reported users willingly using technologies despite their creepiness, e.g.~\cite{phelan_its_2016}. In fact, in Koelle et al.'s~\cite{koelle_social_2020} review of social acceptability research, the WEAR scale is the only mention of creepiness. Thus, related work suggests that creepiness is a distinct concept from acceptability. Here, we investigate creepiness as one's personal perception of an artefact, which is different from acceptability, which is understood as the lack of negative reactions from others~\cite{koelle_social_2020,kelly_wear_2016}. Thus, while acceptability has an inherently social dimension~\cite{toney_social_2003}, we found no previous work that would suggest that creepiness is necessarily social.

We identified only one paper in HCI which investigated the notion of creepiness in an explicit manner. Yip et al.~\cite{yip_laughing_2019} studied perceptions of creepiness in children interacting with technology. They found that the key factors contributing to creepiness were: `deception, lack of control, mimicry, ominous physical appearance, and unpredictability'. They based their inquiry on the socionormative formulation of creepiness by Tene and Polonetsky~\cite{tene_theory_2014}. The results, however, show that creepiness with regard to technology is beyond social norms. The different aspects of creepiness identified by Yip et al. serve as a starting point of our inquiry. The goal of our research is to develop a more structured understanding of these dimensions and a measurement instrument that facilitates comparison in terms of creepiness.

\subsection{Creepy Technologies}
Having established the two domains where the term \emph{creepy} was present, we decided to investigate what other, i.e. non-robot and outside of privacy research, technologies studied in HCI were reported to be creepy. To this end, we conducted a literature review in the ACM Digital Library. We used the query `\texttt{creepy NOT privacy NOT robot}'\footnote{Note that the new ACM DL includes derivative forms of the word, thus, e.g. \emph{creep}, \emph{creeping} and \emph{creepiness} were included as keywords.}, which resulted in 178 papers in SIGCHI sponsored conferences and an additional 9 papers in the ToCHI journal. We then reviewed all the papers and decided to exclude publications which: (1) used the verb \emph{to creep} in a figurative sense or to denote movement, (2) referred to \emph{creep} as a term in materials science, (3) discussed \emph{feature creep}---a phenomenon in software development and (4) used the term \emph{creepy} as part of a citation from prior work. This filtering process yielded 31 papers, which we then open-coded to identify key domains where research reported creepy technologies. While the full results of the review are beyond the scope of this paper, we report here the main areas which we identified with selected examples.

Unsurprisingly, the largest group of papers consisted of papers where the design intention was to make the user feel a certain unease. Exploration through provocative art pieces~\cite{oozu_escaping_2017,lee_tea_2017,kang_scale_2014} or unconventional artefacts~\cite{simbelis_metaphone_2014,paredes_fiat-lux_2016,annett_living_2016} was a prevalent theme in the reviewed corpus. The reported research illustrates how creepiness is an aspect of interactive technologies which designers explicitly consider, thus showing a need for a deeper understanding of the concept. We note that all the artefacts in this group featured differing levels of ambiguity~\cite{gaver_ambiguity_2003}. This suggests that creepiness is connected to not precisely knowing the nature of the artefact. In a similar way, unconventional audio interactions~\cite{rogers_vanishing_2018,kayali_playful_2017} led to not knowing what to expect from a technology and thus experiencing creepiness. For many of these interactive technologies, creepiness may not necessarily be a negative property.

Creepiness when using mediated touch~\cite{pallarino_feeling_2016,levesque_enhancing_2011,israr_towards_2018,doucette_sometimes_2013,gooch_yourgloves_2012} also featured highly in the literature. These papers are considered less relevant to the current paper as the users' perception of unease when using mediated touch was previously defined as \emph{disfordance} by Mejia and Yarosh~\cite{mejia_nine-item_2017} and can be measured with a validated scale. In contrast, our aim is to capture the concept of creepiness for a larger class of artefacts, while building on the lessons learnt from previous research.

Earlier research has noted how interacting with technologies that could be assigned agency was also a source of creepiness. Studies describing interactions with voice assistants~\cite{seymour_does_2020,parviainen_experiential_2020} and with autonomous cars~\cite{moore_case_2019,moore_visualizing_2019,gambino_acceptance_2019} found that both were perceived as creepy. These examples show that creepiness can be experienced where artefacts take an assumed social presence and possibly violate norms related to this presence. An example of this kind of creepiness is how someone feels when crossing the road in front of a driverless car~\cite{moore_visualizing_2019}.

Some of the other papers reported that interactive technologies that have direct contact with our bodies can be perceived as creepy. In particular, creepiness has been reported for wearables~\cite{grinter_ears_2002,genaro_motti_understanding_2014} and technologies that use physiological sensing~\cite{merrill_scanning_2018,merrill_sensing_2019,kang_sharedphys_2016,boem_vitalmorph_2017}. These works offer two ways to frame creepiness. First, we see the notion of a certain magical element, i.e. providing insight one should not have as in Merrill et al.'s~\cite{merrill_sensing_2019} where EEG systems were perceived as mind readers. Second, creepiness is also related to a perception of possible harm~\cite{merrill_scanning_2018}.

Experiences of Augmented Reality~\cite{ni_anatonme_2011,kang_armath_2020} were also potential sources of creepiness. Ni et al.~\cite{ni_anatonme_2011} reported on an Augmented Reality system for facilitating communication with physicians. They equated \emph{a creepy feeling} with \emph{emotional discomfort}.

Additionally, we noted how the term \emph{creepy} has often been used by children~\cite{yip_laughing_2019,kayali_playful_2017,kang_sharedphys_2016}. This is explained by child development research, where it has been found that standards of creepiness are formed early in life~\cite{brink_creepiness_2019}. Furthermore, we even found one paper that reported on creepiness in interactions between the users of a makerspace~\cite{toombs_proper_2015} and one describing social media behaviour~\cite{lim_making_2017} (with no direct connection to privacy).

The variety of findings revealed in our literature review demonstrates the need for a shared conceptual understanding of creepiness within the HCI community. We aim to address this gap by developing a conceptual model of creepiness in HCI and a complementary measurement instrument.

\subsection{Creepiness outside HCI}
The concept of creepiness has also been studied more broadly in the social sciences. For example, Watt and Gallagher~\cite{watt_case_2017} studied how human faces can be defined as creepy. While their study is not directly relevant to the creepiness of inanimate objects, their findings echoed some of the qualitative evidence in HCI where creepiness is linked to violation of norms and the perceived possibility of harm. McAndrew and Koehnke~\cite{mcandrew_nature_2016} used an online survey to establish that unpredictability was a key factor in creepiness. This finding is a relevant aspect for our work as the potentially creepy technologies in HCI research also contained a certain \emph{je ne sais quoi} element. Given the prevalence of the term and its apparent importance for the evaluation of certain classes of technologies, it would be beneficial for HCI to develop a structured understanding of \emph{creepiness} and the means to evaluate if interactive technologies are creepy.

The nearest operationalised concept that has been developed for understanding creepiness in an interactive technology context is Langer and K\"{o}nig's~\cite{langer_introducing_2018}  CRoSS scale. This was designed to rate the creepiness of situations and some of the examined situations involved technology. However, Langer and K\"{o}nig attributed creepiness purely to context. In contrast, our investigation focuses solely on creepiness as a property of an interactive technology. By doing so, our objective is to assess creepiness as part of a design process as opposed to the context. 

\section{Focus Groups}
Our literature review revealed that creepiness has been explained in HCI as a multifaceted concept. Moreover, we found little agreement on what qualities of an artefact contributed to it being perceived as creepy. In order to broaden our understanding of creepiness, we conducted a series of ten focus groups in which participants communicated their first impressions of technologies that could be considered creepy. After the first two focus groups, we refined our focus group protocol. In the first two focus groups, we contrasted different interactive technologies to determine which stimuli were perceived as creepy. This informed our remaining eight focus groups where we focused on one particular creepy technology.

\subsection{Participants}
Eight participants ($M=28.4\,y, SD=3.3\,y, 4$ female$, 4$ male) took part in the first two focus groups. Occupations included business analysts, research associates, students, journalists, project managers and physiotherapists. All participants either had a master's ($37.5\%$) or a bachelor's ($62.5\%$) degree. Participation was voluntary and compensated by 10 Euros.

For the following eight focus groups ($N=24$), we targeted a broader age distribution in order to obtain a more heterogeneous view of the concept of creepiness. Our participant group included older ($M=76.33\,y, SD=3.01\,y$, 6 female, 6 male) and younger adults ($M=27.00\,y, SD=3.16\,y$, 5 female, 7 male). Participants had diverse occupations such as teachers, designers, public servants, students, engineers in different areas including IT, advertising, human resources, political science and psychology. Most participants had a bachelor's degree ($41.7\%$), followed by a master's degree ($25.0\%$) and completed apprenticeships ($16.7\%$). Participation was voluntary and compensated by 10 Euros. See Table \ref{tab:focus_groups_participants} for an detailed overview of the participants.

\begin{table}[htb]
	\centering
	\caption{Participant information for the second set of focus groups, categorised by age group.}
	\begin{tabular}{lllll} 
		\toprule
		\textbf{Variables} & \multicolumn{2}{l}{\textbf{Younger adults}}& \multicolumn{2}{l}{\textbf{Older adults}}\\
		& N & \% & N & \%\\
		\midrule
		\textbf{Gender} & & & &\\
		\quad \footnotesize Women & 5 & 41.7 & 6 & 50.0\\
		\quad \footnotesize Men & 7 & 58.3 & 6 & 50.0\\
		\midrule
		\textbf{Education} & & & &\\
		\quad \footnotesize No degree & 1 & 8.3 & 1 & 8.3\\
		\quad \footnotesize Completed apprenticeship & 2 & 16.7 & 0 & 0\\
		\quad \footnotesize Bachelor's degree, equivalent & 5 & 41.7 & 1 & 8.3\\
		\quad \footnotesize Master's degree, equivalent & 3 & 25.0 & 9 & 75.0\\
		\quad \footnotesize Not specified, equivalent & 1 & 8.3 & 1 & 8.3\\
		\midrule
		\textbf{Work sector} & & & &\\
		\quad \footnotesize (Producing) industry & 1 & 8.3 & 3 & 25.0\\
		\quad \footnotesize Service sector & 3 & 25.0 & 6 & 50.0\\
		\quad \footnotesize Public sector, equivalent & 4 & 33.3 & 0 & 0\\
		\quad \footnotesize Education, equivalent & 0 & 0 & 2 & 16.7\\
		\quad \footnotesize Social services, equivalent & 1 & 8.3 & 1 & 8.3\\
		\quad \footnotesize Undergoing training, equivalent & 3 & 25.0 & 0 & 0\\
		\bottomrule
	\end{tabular}
	\label{tab:focus_groups_participants}
\end{table}

\subsection{Procedure}
The focus groups were divided into an exploratory phase ($25\,min$) and a follow-up group interview ($20\,min$) moderated by one of the experimenters and accompanied by two short questionnaires querying demographics, technology adoption and experience with wearable devices.

In the first two focus groups, participants experienced a set of four technologies varying in aesthetics, comfort and perceived trust. We divided the technologies into conventional devices including the consumer products Fitbit Flex 2\footnote{\url{https://www.fitbit.com/de/flex2}} and the Empatica E4\footnote{\url{https://www.empatica.com/research/e4/}}. Additionally, we included two research prototypes; one showing real-time muscle activity (EMG) through attached electrodes and the other a step-counter based on a pressure-sensitive shoe sole that was attached to the participants' shoes. Under the supervision of two experimenters, the participants took turns in trying out all four devices.

In the follow-up group discussion, the moderator inquired about the participants' first impression of the presented devices. Further topics included their level of trust in the technologies, aspects about the interactive technologies that caught their interest and the perceived interaction with the technologies. Lastly, the group discussed potential usage scenarios.

Based on our findings in the first set of focus groups, the EMG-based technology triggered the most creepy feelings amongst the participants. Thus, we decided to elect the EMG-based device as stimulus for further inquiries.
Three participants took part in each of the eight focus groups after they had been introduced to the EMG prototype. After the introduction and an exploration phase, which lasted $10$ minutes, the focus group started. Again, the focus groups followed a semi-structured protocol and lasted $20\,min$.
We inquired about general perception and interaction with the device. Furthermore, we explored concerns, fears and desires the participants have when interacting with unknown technology. Following a ladder interview approach~\cite{gutman1982means}, we paid special attention to adjectives associated with creepiness, such as creepy, unpleasant, strange, threatening, frightening. In such cases, the moderators explored the topic further. The adjectives were adapted from related work~\cite{langer_introducing_2018,mcandrew_nature_2016,ni_anatonme_2011,watt_case_2017,yip_laughing_2019} with the help of the Oxford Thesaurus of English~\cite{thesaurus}.

\subsection{Data analysis}
All focus group recordings were transcribed verbatim. Then, four authors of the paper open-coded a sample of 20\% of the material and conducted a discussion to establish an initial coding tree in line with Blandford et al.~\cite{blandford2016qualitative}. We further used the factors contributing to creepiness from our literature review as sensitising concepts~\cite{saldana2015coding} in the analysis. The remaining data was distributed equally amongst the four coders. We then iteratively refined the coding of the data that resulted in the construction of three core themes that described the facets of creepiness---as reported by the focus group participants. These themes, together with the insights from our literature review form the foundation of our conceptual model of creepiness.

\section{A Conceptual Model of Creepiness}
\begin{figure*}
    \centering
    \includegraphics[width=.8\textwidth]{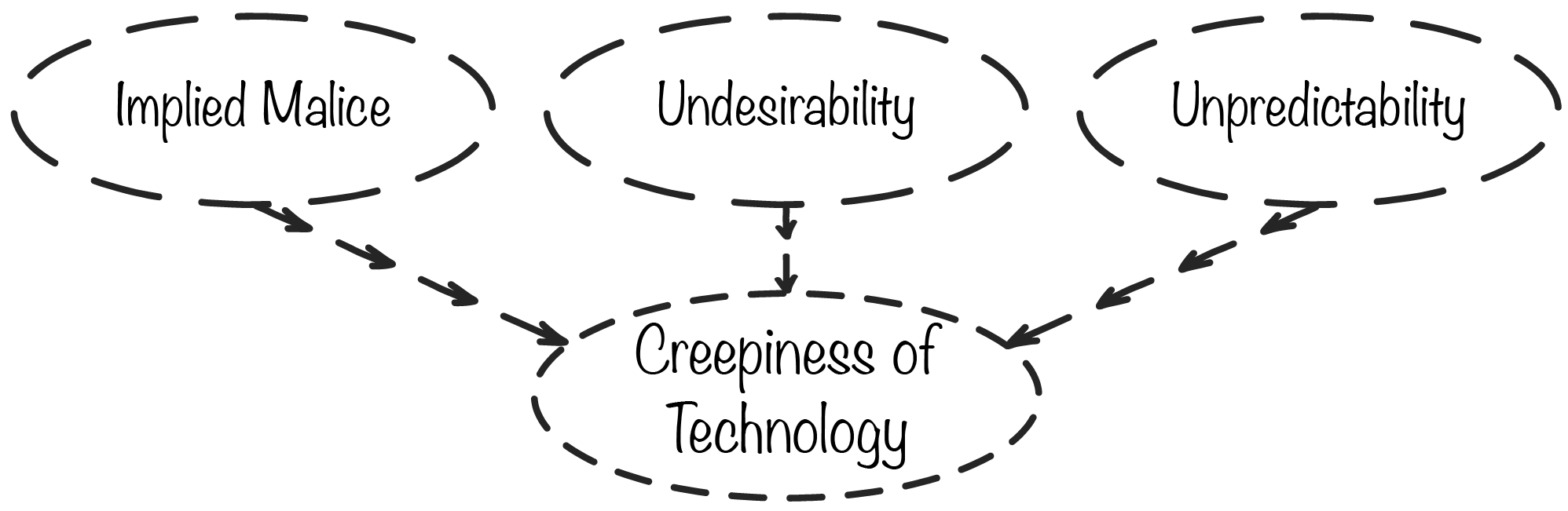}
    \caption{A Conceptual Model of Creepiness. We built the model based on our literature review and qualitative data gathered in the focus groups. The primary task of the model is to inform our design of a scale for measuring creepiness of technology.}
    \label{fig:t-model}
\end{figure*}

Our model of creepiness consists of three dimensions which describe creepy experiences derived from our observations of previous research and the accounts of creepiness derived from our focus groups. The model consists of three dimensions: implied malice, undesirability and unpredictability. The model is shown in Figure~\ref{fig:t-model}.

\subsection{Implied Malice}
This dimension describes perceived bad intentions communicated through the design of a creepy technology. This represents a generalisation of the understanding of creepy as violating privacy. In our model, `implied malice' (\textit{`intention or desire to do evil or cause injury'}~\cite{oed}) described the perceived potential of an interactive artefact that violates principles, which are important to the user.  This dimension is based primarily on reports of creepiness in privacy research and interactions with autonomous systems. Violating the user's value systems was also discussed in the focus groups:

\begin{quote}
   P8: (...) something completely new. Yes, I'm so scared, okay, somehow it's too much (...). For example, when Alexa came on the market, I thought it was super creepy. I still think that it is scary and there are friends of mine and I sometimes think that someone is listening in, for example. (...) I wouldn't buy that myself.
\end{quote}

\subsection{Undesirability}
Undesirability in our model refers to users perceiving the interactive technology as a non sequitur; a feeling of unease caused by the interactive artefact being out of context. The term `undesirability' highlights the feeling of unease inherent to the artefact, which can be due to a variety of factors such as social context or aesthetic appearance. This dimension is based on McAndrew and Koehnke's~\cite{mcandrew_nature_2016} research, adapted to the creepiness of inanimate objects. Focus group participants reflected on negative social consequences of using the technologies with which they were interacting:
\begin{quote}
    P1: But I find it a bit creepy. Imagine you see a person with it (...). P2: I would mainly be worried.
\end{quote}

Undesirability can also imply that the design aesthetic of the artefact does not match the environment in which it is presented to the user. This dimension of creepiness is evident in the provocative technologies discussed above. This was mentioned by one focus group participant who discussed how the aesthetic of the technology did not match its context of use:
\begin{quote}
    P7: Yes. So there must be some serious reason why I have something like that. P9: So when you use something like that; you mentioned suitability for everyday use, [the way this looks] you can't just walk around with it.
\end{quote}

\subsection{Unpredictability}
In our model, we use the term `unpredictability' to denote the negative feelings connected to users not being able to anticipate the interactive technologies’ actions and/or exhibit a desired level of control. In our model, `unpredictability' refers to the inability of the user to immediately operate and understand the device. Control was a key dimension in Yip et al.'s~\cite{yip_laughing_2019} work. Other works demonstrated that a perceived lack of control may lead to a perception of threat~\cite{paredes_fiat-lux_2016}. This dimension also covers the feelings elicited by users not knowing the intended use of an artefact, as discussed by Oozu et al.~\cite{oozu_escaping_2017}. Unpredictability was also a concern for the focus group participants:

\begin{quote}
    P12: For example, I always think about the question, okay, what can the device actually do - the devices of today can do more and more. And you can no longer estimate the [functional] range, (...) I don't even know what it can do.
\end{quote}

\section{The \scale{}}
Having proposed an understanding of the creepiness of interactive artefacts, our next goal is to understand how to ascertain how creepy a given system is. To this end, we decided to build a structured questionnaire. A validated questionnaire would allow designers and researchers to compare artefacts in terms of creepiness levels. Furthermore, through choosing scale items, we could gain additional insight into how users understand creepiness.

We used a structured process to develop our scale, based on the methods recommended by Boateng et al.~\cite{boateng_best_2018}. Given the lack of local standards in HCI for developing questionnaires, our method decisions were also influenced by Mejia and Yarosh's~\cite{mejia_nine-item_2017} work on a related questionnaire designed specifically for use in HCI. 

\subsection{Initial Scale Items}
Four researchers participated in generating initial items for the scale. Researchers first worked independently, creating items based on related work in the three dimensions and quotes from focus group participants. We then conducted a coordination meeting where all the generated items were merged and discussed. After removing duplicates and near-duplicates, we obtained an initial list of 47 items.

\subsection{Expert Feedback}
For the next step, we contacted four experts to provide their feedback on the list of possible scale items. We chose a diverse set of experts to gather broad feedback. The experts were a professor in user modelling, a researcher in machine learning, a researcher in psychology and a user experience lead at a major software company. They provided feedback through commenting on existing items and suggesting new items. Having obtained the feedback, we built a table where we identified problematic items and discussed possible new additions. This process resulted in a final list of 61 items.

\subsection{Survey}
In the next stage of our process, we designed an online survey using the Qualtrics platform to gather data from participants to perform exploratory factor analysis and item reduction. Boateng et al.~\cite{boateng_best_2018}, referring to Comrey~\cite{comrey_factor-analytic_1988}, recommends a sample size of a minimum of 200 participants for studies of this kind and we applied this guideline.

\subsubsection{Participants}
We recruited a total number of $n=209$ participants, which corresponds to the guidelines proposed by Comrey~\cite{comrey_factor-analytic_1988}. The participants were recruited over the Amazon Mechanical Turk Service (MTurk) and reimbursed with 1\$\footnote{\label{footnote:mturk}We used the Qualtrics survey duration estimate, rewarded at a rate approved by the institution of the first author. Based on median completion time, the remuneration was provided at a rate of USD 14 per hour.}. Out of these participants, 109 resided in the European Economic Area and 100 lived in the USA. We informed all participants that study participation was voluntary and if they felt uncomfortable, they could leave at any point. We also informed them that the data collected would be in anonymised form. The survey was conducted online and could be completed in 15 minutes. The average age of the participants was $36\,y\, (SD = 10.6\,y)$ with 33\% identifying as female, 66\% as male and one preferring not to fill in their gender. We asked all participants about their demographics and to fill out a technology adaption scale before the survey.

\subsubsection{Survey content}
In order to evaluate the informative value of our items, we selected four research prototypes in accordance with our model of creepiness. Two prototypes include attaching technology to a user's skin. While one explores opportunities for crafting on-skin interfacing using woven materials~\cite{sun20SecondSkin}, the other looks at the user's hand as a part of an on-skin printed circuit board~\cite{kao18SkinWire}. The other works include prototypes from the domain of mobile devices: a Finger-Navi~\cite{tobita17FingerNavi} integrating the smartphone with a physical finger; and hygiene devices: a teleoperated bottom wiper~\cite{hamada15BottomWiper}.

Each participant in the survey was randomly given a short description and a representative image of exactly one prototype. Afterwards, we asked them how much they agreed with each item of our final list about the presented technology on a 7-item Likert scale (strongly agree to strongly disagree).

\subsection{Exploratory Factor Analysis}
We conducted factor analysis on the survey data collected, using a varimax rotation, thus replicating the method by Mejia and Yarosh~\cite{mejia_nine-item_2017}. We expected the factors to be orthogonal in light of the lack of an established model of creepiness. We chose to perform an orthogonal rotation as the qualitative data suggested that creepiness could be a result of different independent qualities. Further, the different sources of creepiness present in related work suggest an independent relationship~\cite{devellis2016scale}. We used parallel analysis and scree plots to determine the optimal number of factors. The examination of the scree plot suggested an optimal solution with three factors. We then began the process of reducing the number of items. First, we removed all loadings below $0.30$~\cite{boateng_best_2018}. We then removed the items which loaded on multiple factors. This item list was further refined by iteratively removing low loading items and optimising for inter-item reliability. We computed current and theoretical Cronbach's alpha coefficients. Our goal was to create a final scale to be as short as possible for practical reasons---so that it could be deployed by others---be they in industry, academia, government or other---to be able to rapidly compare interactive technologies. The resulting structure consisted of two items loading on one factor and three items loading on the other two factors. We made the non-obvious decision to only use two items for one of the factors as the items loading on that factor were highly correlated and relatively uncorrelated with other items. Worthington and Whittaker~\cite{worthington2006scale} note that two-item constructs are allowable in such cases. While the uneven number of items is not desirable (due to more complicated scoring), this theoretical scale structure offered the best performance in terms of Cronbach's alpha for the scale, $\alpha = 0.74$, and all subscales. The theoretical factor model fit also presented correct parameters at $TLI=0.96$ and $RMSEA = 0.06$. The theoretical composition of the scale is shown in Table~\ref{tab:scale}. The percentage of variance explained was 67.7\% and item communalities were sufficient according to the guidelines set by Hair et al. [A]. The proposed factor structure matched our conceptual model.

\begin{table*}
\caption{The reduced, eight-item Perceived Technology Creepiness Scale (\scaleshort{}). The reported Cronbach's alphas and factor loadings were calculated using the data from the exploratory survey.}
\centering
\begin{tabular}{ll}
\toprule
\textbf{Subscale/Item    }                                                     & \textbf{Factor Loading} \\
\midrule
\textbf{Implied Malice, $\alpha = 0.83$    }                                   &                \\
\midrule
Q1: I think that the designer of this system had immoral intentions.      & 0.86           \\
Q2: The design of this system is unethical.                               & 0.72           \\
\midrule
\textbf{Undesirability, $\alpha = 0.75$  }                                      &                \\
\midrule
Q3: Using this system in public areas will make other people laugh at me. & 0.77           \\
Q4: I would feel uneasy wearing this system in public.                    & 0.85           \\
Q5: The system looks bizarre to me.                                       & 0.70           \\
\midrule
\textbf{Unpredictability, $\alpha = 0.80$    }                                 &                \\
\midrule
Q6: This system looks as expected. (R)                                    & 0.55           \\
Q7: I don't know what the purpose of the system is.                       & 0.84           \\
Q8: This system has a clear purpose. (R)                                  & 0.78          \\
\bottomrule
\end{tabular}
\label{tab:scale}
\end{table*}
\section{Scale evaluation}
Having built a proposed scale with a theoretical underlying factor model, we proceeded to evaluate the \scaleshort{}. We first conducted Confirmatory Factor Analysis to verify the underlying model. Next, we conducted a series of tests to check the scale's construct validity and reliability.

\subsection{Survey}
\subsubsection{Participants}
We recruited $n=100$ participants over MTurk following the first survey. The reimbursement was 0.8\$\footnoteref{footnote:mturk} and the study was conducted online. The study took 5 minutes to complete. The average age of the participants was $34.4\,y\,(SD = 10.1\,y)$, 30\% identified as female and 70\% as male.

\subsubsection{Survey content} 
\begin{figure}
    \centering
    \includegraphics[width=0.45\textwidth]{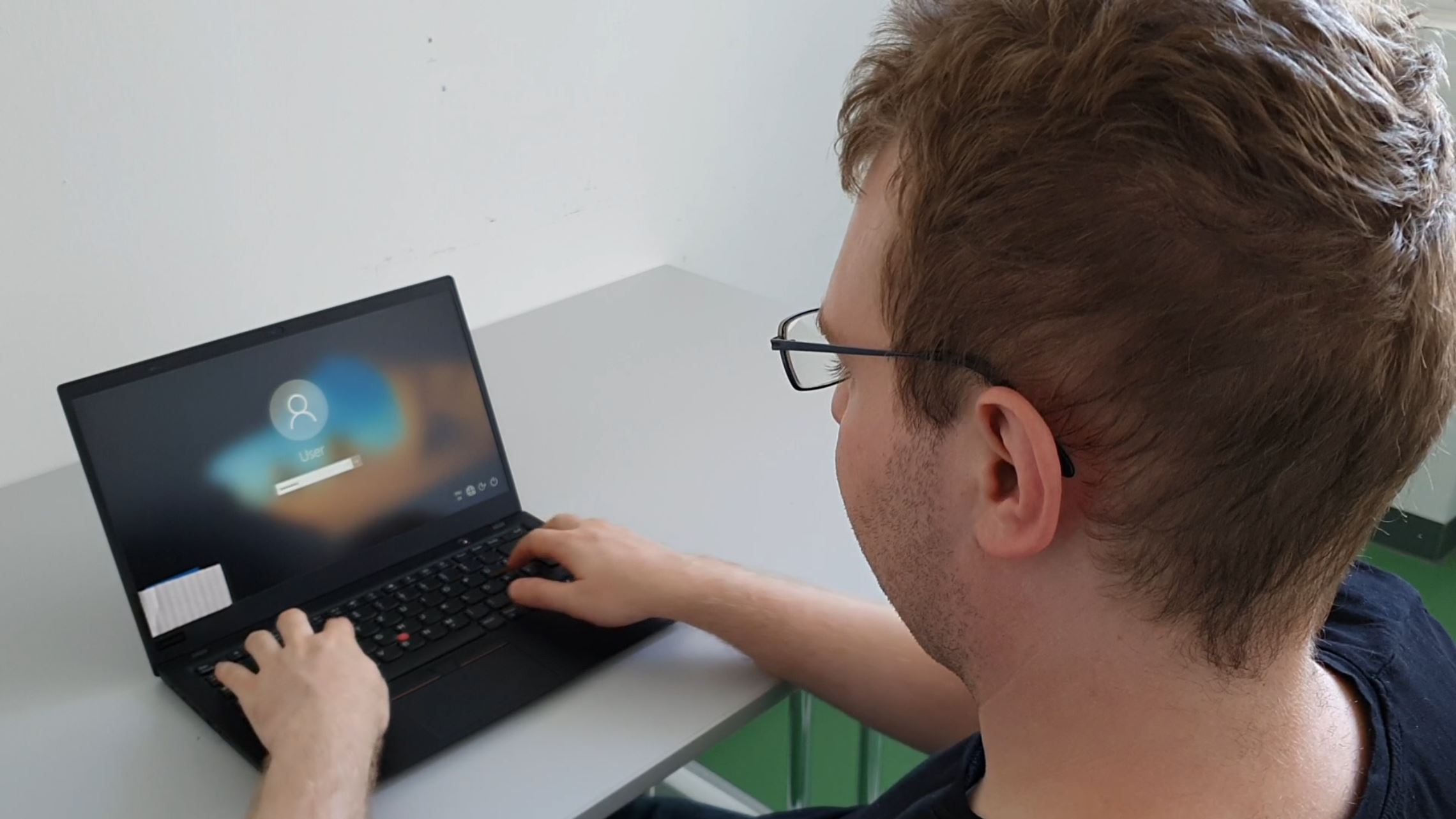}\\[0.4em]
    \includegraphics[width=0.45\textwidth]{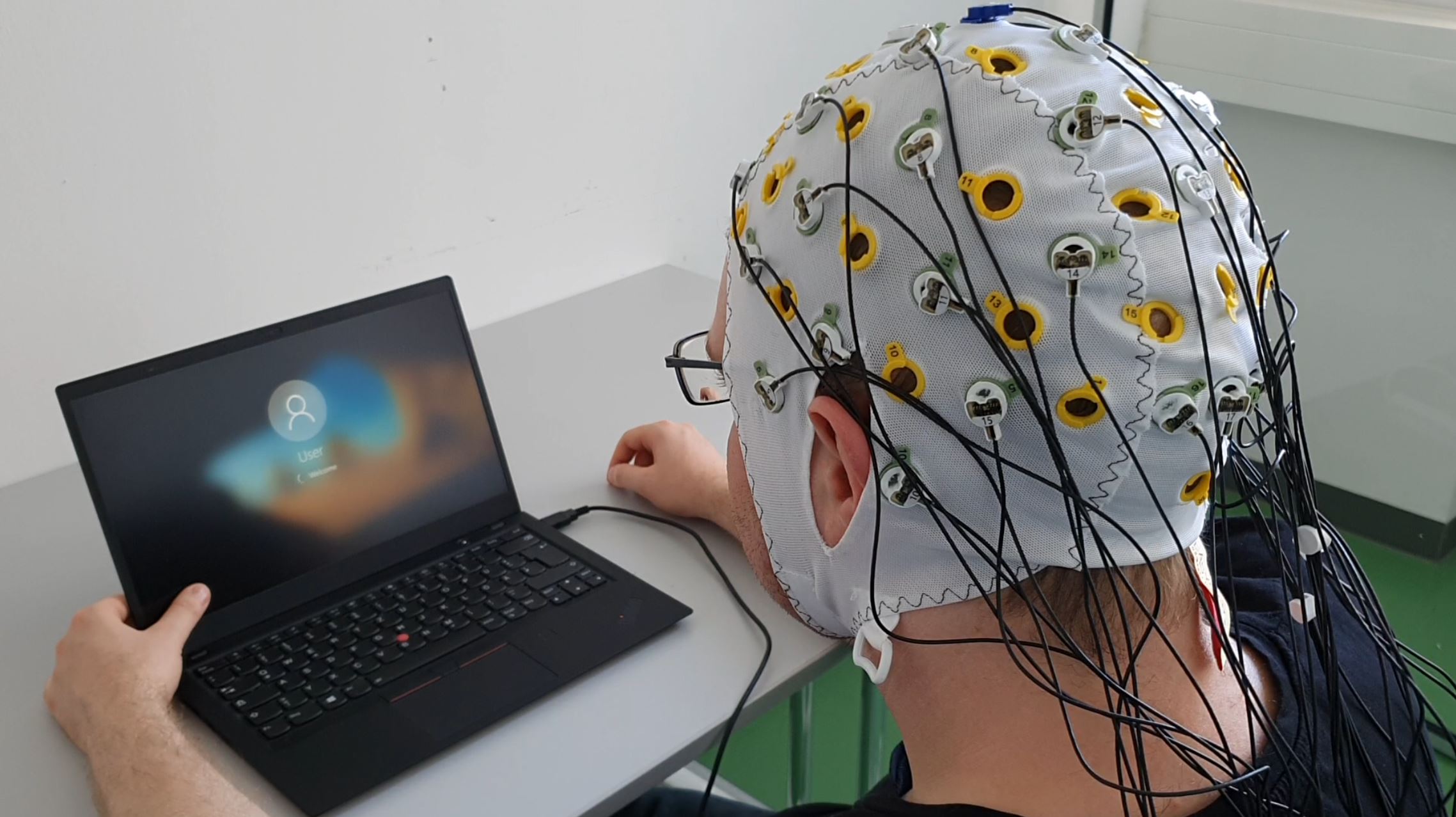}
    \caption{The two different conditions for the scale evaluation. A user entering a password (top) via keyboard and authenticating using their brain waves (bottom).}
    \label{fig:login}
\end{figure}

In order to evaluate the scale, we created two videos of different methods of logging into a computer. One method was typing a password by hand using a keyboard. For the second method, we used an EEG device and participants were told that the user authenticates with their brain waves. Both methods are shown in Figure~\ref{fig:login}.
We randomly presented each participant with one of the two videos. Afterwards, we asked them how much they agreed with each item of our final list about the presented technology on a 7-item Likert scale (strongly agree to strongly disagree).

\subsection{Confirmatory Factor Analysis}
\begin{figure*}
    \centering
    \includegraphics[trim=0 0 0 -120, width=0.8\textwidth]{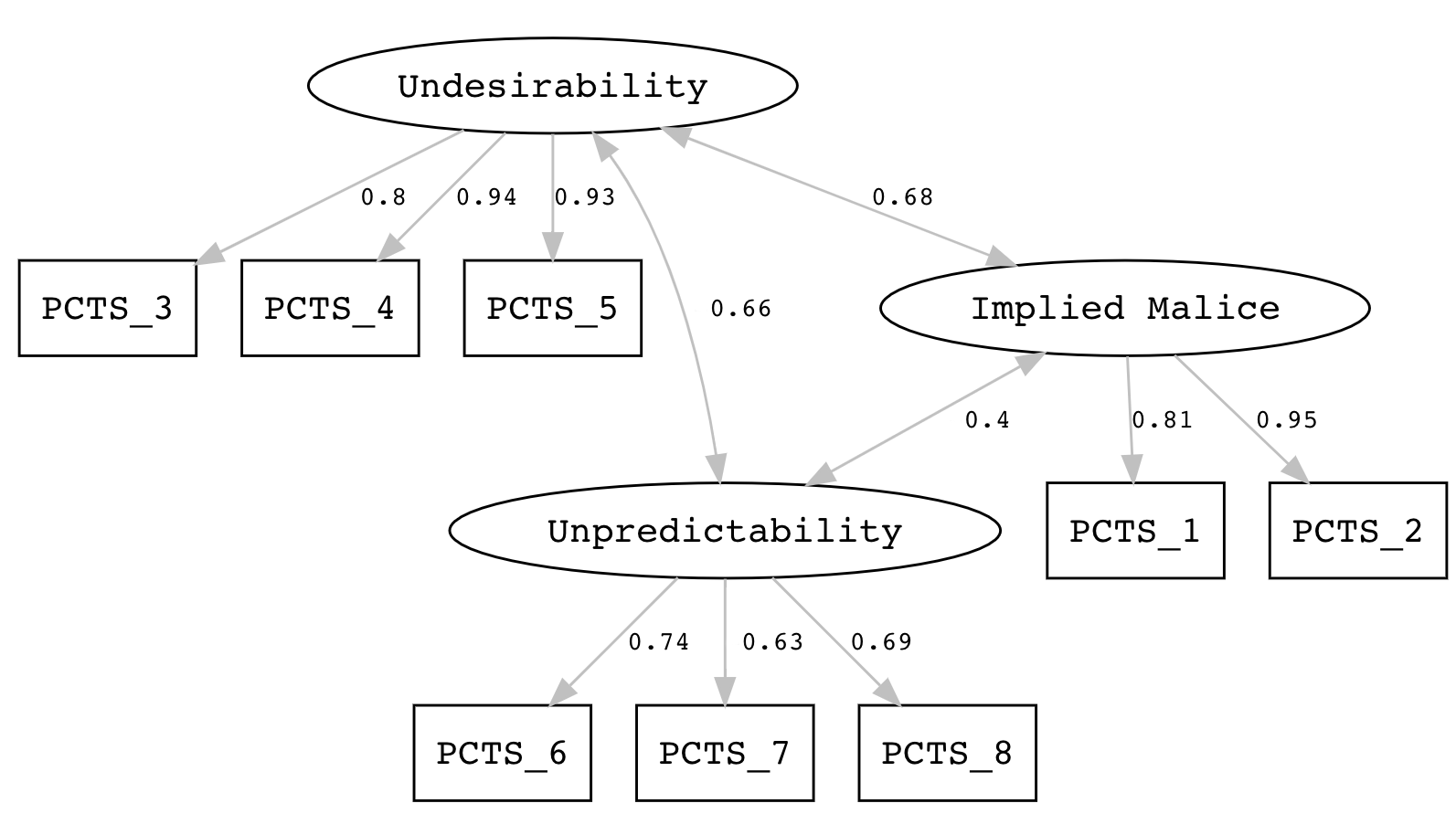}
    \caption{The factor model for \scaleshort{} with the three correlated subscales resulting from confirmatory factor analysis. Note that the graph users inverse scores for reverse-scored items, thus all correlations are positive.}
    \label{fig:CFA}
\end{figure*}

Up to this point, the structure of our scale was only theoretical, i.e. it had not been validated. As a first step in the evaluation of our scale, we conducted Confirmatory Factor Analysis (CFA). This analysis enabled us to conduct a test of dimensionality, which could determine the correctness of our proposed factor model. We used a three-factor model with the latent variables defined as in Table~\ref{tab:scale}. We obtained a fit, which conformed to the required criteria~\cite{boateng_best_2018} with $TLI=1.02$ and $RMSEA < 0.05$. This suggests that the scale was internally consistent. Additionally, the model showed moderate to high correlations between the subscales, showing that the overall construct of creepiness as proposed was valid. The CFA model is shown in Figure~\ref{fig:CFA}.

\begin{table*}
\centering
\caption{Scale evaluation through differentiation by known groups for \scaleshort{}. Non-parametric tests show that logging in via EEG was significantly more creepy than using only the keyboard using the full scale and the subscales. Table reports Bonferroni-corrected p-values.}
\begin{tabular}{lllllll}
\toprule
Scale/Subscale & $M_{Keyboard}$ & $SD_{Keyboard}$ & $M_{EEG}$ & $SD_{EEG}$ & $U$      & $p$      \\
\midrule
\scaleshort{}           & 22.24 & 13.62   & 35.12 &  7.80    & $1978.0$ & $<0.001$ \\
\scaleshort{}-IM        & 7.74  &  5.72   & 9.73  & 4.55    & $1670.5$ & $<0.05$  \\
\scaleshort{}-UD        & 7.52  &  5.37  & 14.80  & 4.02   & $2144.0$ & $<0.001$ \\
\scaleshort{}-UP        & 6.98  &  3.89   & 10.58  & 3.53   & $1931.5$ & $<0.001$\\
\bottomrule
\end{tabular}
\label{tab:groups}
\end{table*}

\subsection{Construct validity}
Next, we examine the construct validity of the \scaleshort{}. We decided to test the scale in two ways. First, we checked if the scale was effectively differentiating between `known groups', i.e. interactive technologies that differ in creepiness. Second, we investigated if the scale was different from possibly related concepts measured in other questionnaires.
\subsubsection{Differentiation by `known groups'}
Boateng et al.~\cite{boateng_best_2018} listed comparison between `known groups' as a method of establishing concept validity. Mejia and Yarosh~\cite{mejia_nine-item_2017} also used this method. In our work, we conducted a comparison between a system known (albeit qualitatively) to be creepy in the literature~\cite{merrill_sensing_2019}---an EEG system, and a conventional solution with which the users were familiar---the keyboard. We hypothesised that logging in with EEG would be significantly more creepy than logging in with only a keyboard. Shapiro-Wilk tests revealed that the samples were not normally distributed. Thus, we applied non parametric statistics. Table~\ref{tab:groups} shows Mann-Whitney U test results for \scaleshort{} and its subscales.

\subsubsection{Discriminant validity}
Discriminant validity refers to how a scale measures concepts that are different from other scales. Given the conceptual model behind building the \scaleshort{}, we wanted to check if creepiness was not simply a reflection of social acceptability or anticipated usability. As such, a comparison is only possible with other validated questionnaires, and so our choice of alternative concepts was limited. So, we decided to investigate if the \scaleshort{} provided measures different from the dimensions of the Technology Acceptance Model, as measured by a highly-cited questionnaire by Park~\cite{park_analysis_2009}. This questionnaire featured a number of factors which could be potential confounding concepts for the \scaleshort{}: perceived ease of use (PE), perceived usefulness (PU), Attitude (AT) and Behavioural intent (BI). Furthermore, we ensured that the \scaleshort{} measured properties of the artefact and not the user's personality in terms of attitudes towards technology. To this end, we compared \scaleshort{} scores with McKnight et al.'s Propensity to Trust in General Technology (PTT, \cite{mcknight_trust_2011}). We computed Spearman correlations between the different scales. The results, shown in Table~\ref{tab:corr}, show at most medium to low correlations between the \scaleshort{} or its subscales and the other measurement instruments. The medium correlation suggests that some of the instruments may be measuring contextual factors related to the \scaleshort{}. There might also be an overlap between some of the dimensions of the \scaleshort{} and the dimensions developed by Park~\cite{park_analysis_2009}, which does not impact the overall scores of the \scaleshort{}. These results show that the \scaleshort{} is a novel concept and validates \scaleshort{}'s underlying model.

\begin{table*}
\centering
\caption{Spearman correlations between the \scaleshort{}, its subscales and potential alternative questionnaires that could measure creepiness. Significant correlation tests are marked with an asterisk.}
\begin{tabular}{rllllllll}
  \hline
& \scaleshort{} & \scaleshort{}-IM & \scaleshort{}-UD & \scaleshort{}-UP & PTT & PE & PU & AT \\ 
  \hline
  \scaleshort{}-IM &  0.85* &  &  &  &  &  &  &  \\ 
  \scaleshort{}-UD &  0.89* &  0.66** &  &  &  &  &  &  \\ 
  \scaleshort{}-UP &  0.70* &  0.43** &  0.52* &  &  &  &  &  \\ 
  PTT & -0.22*    & -0.24*    & -0.09     & -0.32*  &  &  &  &  \\ 
  PE & -0.38* & -0.31*   & -0.44* & -0.67* &  0.34*  &  &  &  \\ 
  PU &  0.03     &  0.22*    & -0.03     & -0.19     &  0.20*    &  0.39* &  &  \\ 
  AT & -0.38* & -0.30*   & -0.48* & -0.59* &  0.31*   &  0.72* &  0.55** &  \\ 
  BI & -0.42* & -0.24*    & -0.45* & -0.50* &  0.29*   &  0.70* &  0.54* &  0.86* \\ 
   \hline
\end{tabular}
\label{tab:corr}
\end{table*}

\subsection{Test-Retest Reliability}
As a final evaluation of the scale, we tested its temporal stability, i.e. whether the scale can produce reliable results at different time points. To this end, we administered the \scaleshort{} to a group of $n=20$ participants, aged $M=29.15\,y, SD=3.12\,y$, 15 male and 5 female, twice, with a minimum 14-day break in between the studies.

There is a lack of consensus in the literature about how long the time between the two surveys should be. We decided to replicate Mejia and Yarosh's~\cite{mejia_nine-item_2017} approach, albeit with a larger participant sample. In an online survey, the participants were asked to rate an artefact previously qualitatively reported to evoke feelings of creepiness---the HEXBUG~\cite{somanath_integrating_2013}. Contrary to the previous studies, we used snowball sampling and social media posts to recruit the participants. This allowed us to ensure we could reach participants effectively to ask them to conduct the survey for the second time.

Boateng et al.~\cite{boateng_best_2018} list a number of ways to assess test-retest reliability. We chose to compute the intra-class correlation coefficient~\cite{koo_guideline_2016} to investigate the relationship between the two measurement moments. There was a high reliability with $\kappa=0.82$, $p<.001$. The 95\% confidence interval ranged from $0.65$ to $0.91$, indicating that \scaleshort{} exhibited moderate to excellent test-retest reliability~\cite{cohen_statistical_2013}.  The results show that the \scaleshort{} is temporally stable and can be administered at different times or used for between-groups or repeated-measures designs.

\section{Discussion}
In this section, we provide the necessary details for administering the \scaleshort{} as well as information on how to analyse the results. In addition, we discuss the limitations of our approach and possibilities for further development.

\subsection{Scoring and Analysis}
The \scaleshort{} is scored on a seven-point Likert scale from Strongly Agree (7) to Strongly Disagree (1). Items 6 and 8 in the scale are reverse-scored. Our scale has a $2+3+3$ item composition, which makes computing scores less than trivial. One possible solution to optimally balance the items would be extracting the weights from the factor models developed and then assigning those weights to the three subscales. However, the review by Boateng et al.~\cite{boateng_best_2018} concluded that using weighted averages is unlikely to improve the performance of questionnaires. Consequently, we suggest assigning equal weights to the individual subscales. The \scaleshort{}-IM score is thus calculated as the sum of the two items multiplied by $1.5$.
Consequently, the \scaleshort{} is scored as (reverse-scored items are marked with the subscript $R$):
\begin{align*}
    \scaleshort{}&=\scaleshort{}_{IM}+\scaleshort{}_{UD}+\scaleshort{}_{UP}\\
    \; where \; \scaleshort{}_{IM}&=(Q1+Q2)\times 1.5 \notag\\
    and \; \scaleshort{}_{UD}&=Q3+Q4+Q5 \; \\
    and \; \scaleshort{}_{UP}&=Q6_{R}+Q7+Q8_{R}
\end{align*}
Thus, the lowest score on the scale is 9 and the highest is 63. Higher scores indicate that the interactive artefact evokes stronger feelings of creepiness. This scoring offers a transparent and actionable way for designers and researchers to use the \scaleshort{}. The evaluation of the scale presented in this paper suggests that conducting null-hypothesis testing using \scaleshort{} scores and its subscales score is permitted. We recommend checking the normality of the data and possibly using non-parametric statistics when conducting experiments using the scale.

\subsection{Guidelines for using the \scaleshort{}}
The robust structure of the \scaleshort{} allows using it for different study designs within HCI research as well as for quick assessments of research prototypes. We particularly recommend using the \scaleshort{} in early stages of the design process. The scale is relatively short, easy to use and can offer rapid feedback. This can help in identifying artefacts early in the design process which appear creepy to users and help raise awareness of how to steer the design process in a more desirable direction.

We developed the \scaleshort{} primarily to capture users' first impressions of experiencing an artefact. Consequently, the scale is particularly suited to studying initial encounters with technologies, discovering new (physiological) sensing modalities or interacting with previously unknown aesthetic forms. The scale examines an aspect of user experience beyond acceptance and usability, as indicated by our results. We recommend using the \scaleshort{} to identify features of artefacts that may be creepy early in the design process. Using the \scaleshort{} enables for effectively managing first impressions of technologies and ensuring that the technology does not intimidate the user to a point where they are unwilling to verify its usability. \scaleshort{} can facilitate rapid selection of solutions at the prototype generation phase in the user-centred design process.

In this context, future users of the scale should recognise that creepiness is not necessarily a negative aspect of technologies. The provocative or intentionally ambiguous technologies discussed in this paper may use creepiness for the benefit of users. This suggests that creepiness is a highly contextualised concept. Hence, our scale is best used when comparing between different technologies within the same context. Consequently, future users of the \scaleshort{} should carefully control the context in which the participants are introduced to the artefacts studied in order to avoid bias.

The psychometric properties of the \scaleshort{} indicate that the scale can be used for between- and within-subject studies and for repeated-measures designs. If particular aspects of a given technology are of importance, e.g. its ethical underpinnings, the subscales of the \scaleshort{} can also be analysed. However, we recommend that the use of \scaleshort{} be accompanied by pre-studies and rich qualitative data gathering. Potential users of the \scaleshort{} should be sure that the technology studied may evoke creepiness in the understanding of the \scaleshort{}, i.e. an innate, hard to define feeling of unease. Here, the \scaleshort{} may be used to help the designer reflect on the potential creepiness. Alternative, more detailed questionnaires can be used if the main concern about `creepiness' is indeed a question of privacy encroachment~\cite{chignell_privacy_2003} or social acceptability~\cite{rico_gestures_2009}. Our results show that the PCTS measures a concept different than usability or social acceptability. Thus, our scale broadens the apparatus available to HCI researchers in quantitatively understanding impressions of new technologies.

\subsection{Limitations}
We recognise that the development and possible use of the \scaleshort{} is prone to certain limitations. First, we made the decision to focus the development of the scale on assessing the initial impressions of technologies. This implies that the usefulness of the \scaleshort{} for long-term studies is unknown. We envision that the scale could be used to measure how users gradually get more acquainted with a technology and their perception of it changes. This would be particularly relevant for better understanding interactive technologies with which users develop long-term relationships, e.g. voice assistants~\cite{purington_alexa_2017}. In future research, we plan to evaluate if the \scaleshort{} can be used effectively beyond first impressions.

While we used a number of recruitment strategies and study methods, we recognise that the development of the \scaleshort{} is biased by the participant sample used. The focus groups which highly influenced our conceptual model of creepiness were conducted solely among residents of Europe. The majority of the participants in the studies which used MTurk recruitment in our work had a Western cultural background. The term \emph{creepy} is a modern English word that is difficult to translate to many languages. Consequently, we note that the \scaleshort{} is most likely only applicable to users with a selected subset of cultural backgrounds. We hope that future research can develop alternative versions of the scale for other cultural contexts.

\section{Conclusion}
In this paper, we presented the development and evaluation of the \scale{} (\scaleshort{}). Based on a literature review and focus groups, we developed a conceptual model for creepiness. We then describe how we constructed, reduced and evaluated the scale. We illustrated the discriminant validity of the scale, its ability to differentiate between known groups and test-retest reliability. Our scale enables designers and researchers to rapidly ascertain possible feelings of unease caused by novel interactive technologies. The \scaleshort{} can be used to conduct rapid comparative studies of novel artefacts, especially ones that exhibit elements of autonomy or feature direct contact with the body.

We designed the \scaleshort{} with the goal of enabling a broader understanding of how current and future technologies make us feel and how to build technologies that do not cause negative emotions in users. We also note that our scale can help in studying possibly provocative artefacts that could foster engagement. We hope that our scale can foster new research avenues into increasing our understandings of creepiness and to enlighten those designing to avoid (or to promote) creepiness by providing them with a creepiness metric they can easily use to  conduct studies of novel technologies.

\begin{acks}
This work was supported through multiple funding schemes. First, we like to acknowledge the support of the Leibniz ScienceCampus Bremen Digital Public Health (\url{https://www.lsc-digital-public-health.de/}), which is jointly funded by the Leibniz Association (W4/2018), the Federal State of Bremen and the Leibniz Institute for Prevention Research and Epidemiology---BIPS. This research is also partially funded by the German Research Foundation (DFG) under Germany´s Excellence Strategy (EXC 2077, University of Bremen) and a Lichtenberg professorship funded by the Volkswagen foundation. This work was financially supported by Utrecht University's Focus Area: Sports and Society and the European Union’s Horizon 2020 Programme under ERCEA grant no. 683008 AMPLIFY.
\end{acks}

\bibliographystyle{ACM-Reference-Format}
\bibliography{Creepy}


\end{document}